\input{aipcheck}
\documentclass[final,numberedheadings]{aipproc}
\usepackage{epsfig,latexsym}
\layoutstyle{6x9}

\newcommand{\agt}{\,\rlap{\lower 3.5 pt \hbox{$\mathchar \sim$}} \raise 1pt
 \hbox {$>$}\,}
\newcommand{\alt}{\,\rlap{\lower 3.5 pt \hbox{$\mathchar \sim$}} \raise 1pt
 \hbox {$<$}\,}

\begin{document}

\title{Charmonium production in two-photon collisions at next-to-leading
order}


\author{Bernd A. Kniehl}{
 address={II. Institut f\"ur Theoretische Physik, Universit\"at Hamburg,
Luruper Chaussee 149, 22761 Hamburg, Germany}
}

\begin{abstract}
We review recent results on the production of prompt charmonium in association
with a hadron jet or a prompt photon in two-photon collisions at
next-to-leading order in the factorization framework of nonrelativistic
quantum chromodynamics.
\end{abstract}

\maketitle

\section{Introduction}

The factorization formalism of nonrelativistic quantum chromodynamics (NRQCD)
\cite{bbl} provides a rigorous theoretical framework for the description of
heavy-quarkonium production and decay that is renormalizable and predictive.
Theoretical predictions are decomposed into sums over products of
short-distance coefficients, which can be calculated perturbatively as
expansions in the strong-coupling constant $\alpha_s$, and long-distance
matrix elements (MEs), which are subject to relative-velocity ($v$) scaling
rules and must be extracted from experiment, and are so organized as double
expansions in $\alpha_s$ and $v$.
This formalism takes into account the complete structure of the
$Q\overline{Q}$ Fock space, which is spanned by the states
$n={}^{2S+1}L_J^{(a)}$ with definite spin $S$, orbital angular momentum
$L$, total angular momentum $J$, and color multiplicity $a=1,8$, and so
predicts the existence of color-octet (CO) processes in nature.

The greatest triumph of the NRQCD factorization formalism was its ability to
correctly describe the cross section of inclusive charmonium hadroproduction
at the Tevatron, which exceeds the color-singlet-model prediction by more
than one order of magnitude.
In order to convincingly establish the phenomenological significance of the
CO processes, it is indispensable to identify them in other kinds of
high-energy experiments as well.
The verification of the NRQCD factorization hypothesis is presently hampered
both from the theoretical and experimental sides.
On the one hand, the theoretical predictions to be compared with existing
experimental data are, apart from very few exceptions, of lowest order (LO)
and thus suffer from considerable uncertainties.
The measurement of charmonium polarization at the Tevatron currently presents
a challenge for NRQCD factorization, but any conclusions are premature in the
absence of a full-fledged next-to-leading-order (NLO) analysis.
It is, therefore, mandatory to calculate the NLO corrections to the
hard-scattering cross sections and to include the composite operators that are
suppressed by higher powers in $v$.
Apart from the usual reduction of the renormalization and factorization scale
dependences, sizeable effects, e.g.\ due to the opening of new partonic
production channels, are expected at NLO.
On the other hand, the experimental errors are still rather sizeable.
The latter are being significantly reduced by HERA~II and run~II at the
Tevatron, and will be dramatically more so by the LHC and hopefully a future
$e^+e^-$ linear collider such as the TeV-Energy Superconducting Linear
Accelerator (TESLA), which is presently being designed and planned at DESY.

Recently, $2\to2$ processes of heavy-quarkonium production were for the first
time studied at NLO in the NRQCD factorization formalism \cite{nlo,nlog}.
Specifically, the production of prompt charmonium, which is produced either
directly or through the decay of heavier charmonia, with finite transverse
momentum ($p_T$) in association with a hadron jet \cite{nlo} or a prompt
photon \cite{nlog} via direct photoproduction in two-photon collisions was
considered.
In this presentation, we review the most important conceptional issues and
phenomenological results of Refs.~\cite{nlo,nlog}, in Sections~\ref{sec:two}
and \ref{sec:three}, respectively.

\section{Conceptional issues}
\label{sec:two}

We focus attention on the process $\gamma\gamma\to J/\psi+X$, where $J/\psi$
is promptly produced at finite value of $p_T$ and $X$ is a purely hadronic
remainder.
Since the incoming photons can interact either directly with the quarks
participating in the hard-scattering process (direct photoproduction) or via
their quark and gluon content (resolved photoproduction), this process
receives contributions from the direct, single-resolved, and double-resolved
channels, which are formally of the same order in the perturbative expansion.
At LO, the bulk of the cross section is due to single-resolved photoproduction,
and the NRQCD prediction \cite{prl} based on the MEs determined from fits
\cite{prl} to Tevatron data nicely agrees with a recent measurement by the
DELPHI Collaboration at LEP2 \cite{delphi}.

Here, we consider direct photoproduction at NLO \cite{nlo}.
At LO, there is only one partonic subprocess, namely
\begin{equation}
\gamma+\gamma\to c\overline{c}[{}^3\!S_1^{(8)}]+g.
\label{eq:lo}
\end{equation}
At NLO, virtual corrections to process~(\ref{eq:lo}) and
$\langle{\cal O}^H[{}^3\!S_1^{(8)}]\rangle$, where
$H=J/\psi,\chi_{cJ},\psi^\prime$, and real corrections to
\begin{eqnarray}
&\gamma+\gamma\to c\overline{c}[n]+g+g,\qquad
&n={}^3\!P_J^{(1)},{}^1\!S_0^{(8)},{}^3\!S_1^{(8)},{}^3\!P_J^{(8)},
\label{eq:gg}\\
&\gamma+\gamma\to c\overline{c}[n]+u+\overline{u},\qquad
&n={}^3\!S_1^{(8)},
\label{eq:uu}\\
&\gamma+\gamma\to c\overline{c}[n]+q+\overline{q},\qquad
&n={}^1\!S_0^{(8)},{}^3\!S_1^{(8)},{}^3\!P_J^{(8)},
\label{eq:qq}
\end{eqnarray}
where $u$ and $\overline{u}$ denote the Faddeev-Popov ghosts of the gluon,
contribute.

The virtual corrections to process~(\ref{eq:lo}) receive contributions from
self-energy, triangle, box, and pentagon diagrams.
The self-energy and triangle diagrams are in general ultraviolet (UV)
divergent;
the triangle, box, and pentagon diagrams are in general infrared (IR)
divergent;
and the pentagon diagrams without three-gluon vertex also contain Coulomb
divergences.
As for the light-quark loops, the triangle diagrams vanish by Furry's theorem,
while the box diagrams form a finite subset.
The virtual corrections to $\langle{\cal O}^H[{}^3\!S_1^{(8)}]\rangle$ also
produce UV, IR, and Coulomb divergences.
The UV and IR divergences are extracted using dimensional regularization in
$d=4-2\epsilon$ space-time dimensions, leading to poles in $\epsilon_{\rm UV}$
and $\epsilon_{\rm IR}$, respectively, while the Coulomb singularities are
regularized by a small value of $v$.
The UV divergences are removed by the renormalization of
$\langle{\cal O}^H[{}^3\!S_1^{(8)}]\rangle$, $\alpha_s$,
the charm-quark mass and field, and the gluon field, which is performed in the
modified minimal-subtraction ($\overline{\rm MS}$) scheme for the former two
quantities, rendering them dependent on the renormalization scales $\lambda$
and $\mu$, respectively, and in the on-mass-shell scheme for the residual
three quantities.
The IR divergences cancel among the virtual and real corrections,
the wave-function renormalizations, and
$\langle{\cal O}^H[{}^3\!S_1^{(8)}]\rangle$.
The Coulomb divergences cancel between the virtual corrections and
$\langle{\cal O}^H[{}^3\!S_1^{(8)}]\rangle$.

The real corrections are plagued by IR divergences, which come as collinear
divergences from the initial state and collinear and/or soft ones from the
final state.
They are identified by appropriately slicing the three-particle phase space
using small parameters $\delta_i$ and $\delta_f$, respectively.
The collinear and/or soft regions of phase space are integrated over
analytically in $d$ dimensions, while the hard region is integrated over
numerically in four dimensions.
The sum of these contributions is, to very good approximation, independent of
$\delta_i$ and $\delta_f$.
The initial-state collinear divergences are factorized at some factorization
scale $M$ and absorbed into the parton density functions (PDFs) of the $q$ and
$\overline{q}$ quarks inside the resolved photon.
The $M$ dependence thus introduced is approximately compensated by the LO
single-resolved contribution.

Combining the contributions arising from the virtual corrections (vi), the
parameter and wave-function renormalization (ct), the operator redefinition
(op), the initial-state (is) and final-state (fs) collinear configurations,
the soft-gluon radiation (so), and the hard-parton emission (ha) as
\begin{eqnarray}
\lefteqn{d\sigma(\mu,\lambda,M)=
d\sigma_0(\mu,\lambda)[1
+\delta_{\rm vi}(\mu;\epsilon_{\rm UV},\epsilon_{\rm IR},v)
+\delta_{\rm ct}(\mu;\epsilon_{\rm UV},\epsilon_{\rm IR})
+\delta_{\rm op}(\mu,\lambda;\epsilon_{\rm IR},v)}\nonumber\\
&&{}+\delta_{\rm fs}(\mu;\epsilon_{\rm IR},\delta_f)]
+d\sigma_{\rm is}(\mu,\lambda,M;\delta_i)
+d\sigma_{\rm so}(\mu,\lambda;\epsilon_{\rm IR},\delta_f)
+d\sigma_{\rm ha}(\mu,\lambda;\delta_i,\delta_f),\quad
\label{eq:sum}
\end{eqnarray}
the regulators $\epsilon_{\rm UV}$, $\epsilon_{\rm IR}$, $v$, $\delta_i$, and
$\delta_f$ drop out and the $\mu$ and $\lambda$ dependences formally cancel up
to terms beyond NLO, while the $M$ dependence is unscreened at NLO.

\section{Phenomenological results}
\label{sec:three}

We consider two-photon collisions at TESLA operating at a center-of-mass
energy of 500~GeV, where the photons arise from electromagnetic initial-state
bremsstrahlung, with antitagging angle $\theta_{\rm max}=25$~mrad, and
beamstrahlung, with effective beamstrahlung parameter $\Upsilon=0.053$.
The $J/\psi$, $\chi_{cJ}$, and $\psi^\prime$ MEs are adopted from
Ref.~\cite{bkl} and the photon PDFs from Ref.~\cite{grs}.

\begin{figure}[ht]
\begin{tabular}{cc}
\parbox{0.45\textwidth}{
\epsfig{file=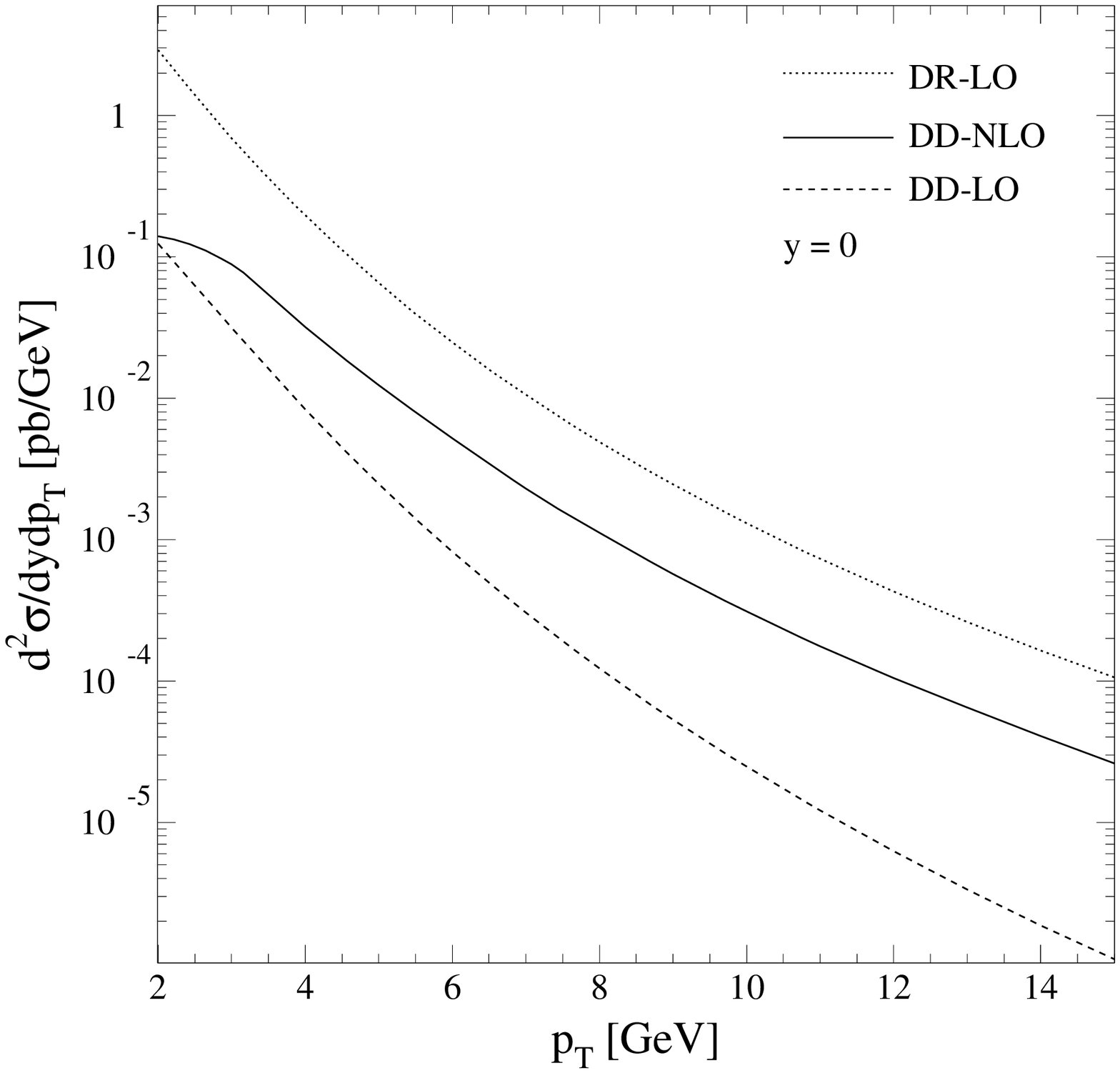,width=0.45\textwidth,%
bbllx=0pt,bblly=0pt,bburx=530pt,bbury=513pt}
}
&
\parbox{0.45\textwidth}{
\epsfig{file=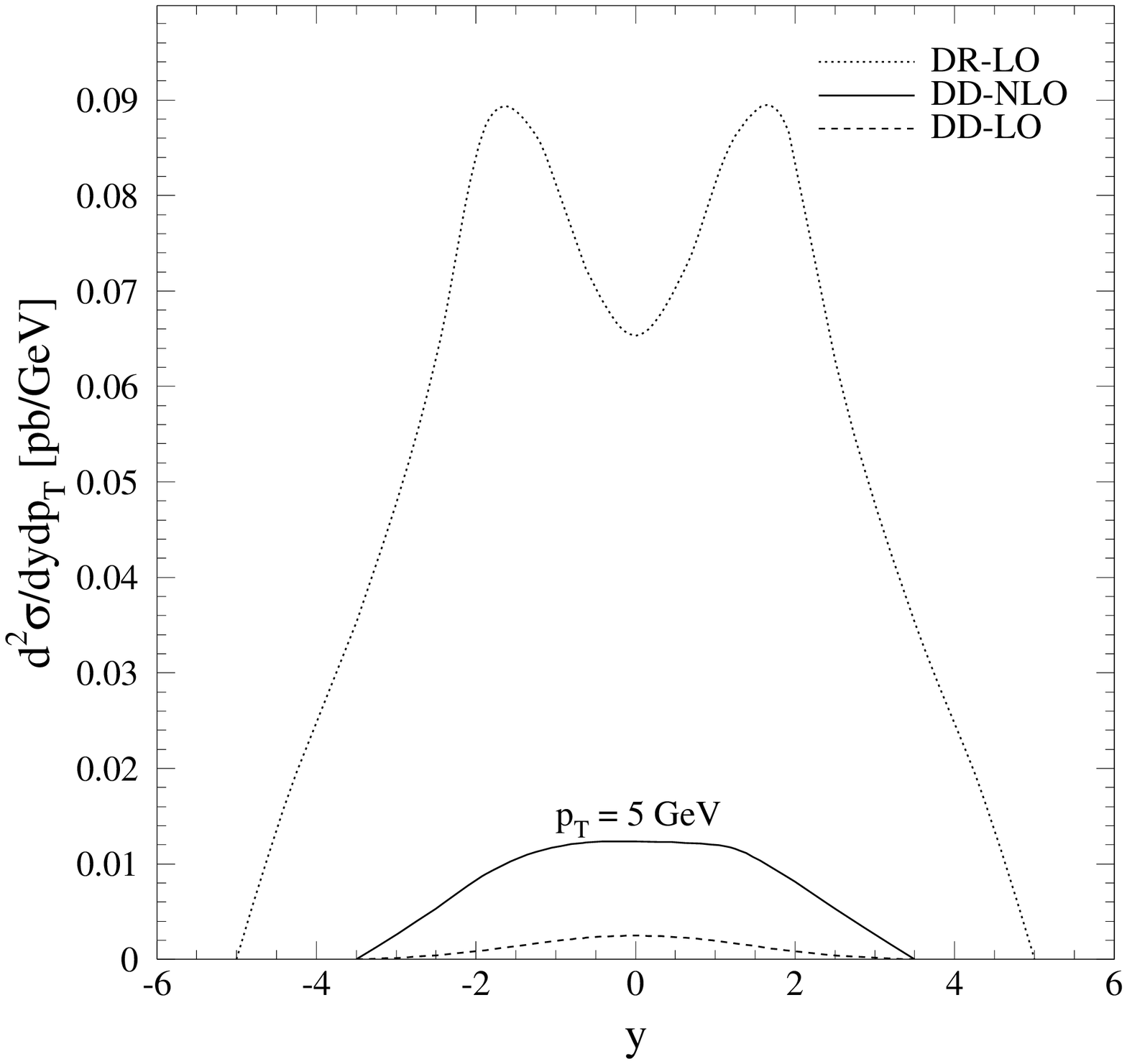,width=0.45\textwidth,%
bbllx=0pt,bblly=0pt,bburx=530pt,bbury=513pt}
}
\end{tabular}
\caption{LO single-resolved (dotted lines), LO direct (dashed lines), and NLO
direct (solid lines) contributions to $d^2\sigma/dp_Tdy$ (a) for $y=0$ as a
function of $p_T$ and (b) for $p_T=5$~GeV as a function of $y$.}
\label{fig:xs}
\end{figure}

In Fig.~\ref{fig:xs}, we study $d^2\sigma/dp_Tdy$ (a) for rapidity $y=0$ as a
function of $p_T$ and (b) for $p_T=5$~GeV as a function of $y$, comparing the
LO (dashed lines) and NLO (solid lines) results of direct photoproduction with
the LO result of single-resolved photoproduction (dotted lines).
From Fig.~\ref{fig:xs}(a), we observe that, with increasing value of $p_T$,
the NLO result of direct photoproduction falls of considerably more slowly
than the LO one.
In fact, the QCD correction ($K$) factor, defined as the NLO to LO ratio,
rapidly increases with $p_T$, exceeding 10 for $p_T\agt10$~GeV.
This feature may be understood by observing that so-called
{\it fragmentation-prone} \cite{jb} partonic subprocesses start to contribute
to direct photoproduction at NLO, while they are absent at LO.
Such subprocesses contain a gluon with small virtuality, $q^2=4m_c^2$, that
splits into a $c\overline{c}$ pair in the Fock state $n={}^3\!S_1^{(8)}$ and
thus generally generate dominant contributions at $p_T\gg2m_c$ due to the
presence of a large gluon propagator.
In single-resolved photoproduction, a fragmentation-prone partonic subprocess
already contributes at LO.
This explains why the solid and dotted curves in Fig.~\ref{fig:xs}(a) run
parallel in the upper $p_T$ range.
At low values of $p_T$, the fragmentation-prone partonic subprocesses do not
matter, and the relative suppression of direct photoproduction is due to the
fact that, at LO, this is a pure CO process.

In the case of $\gamma\gamma\to J/\psi+X_\gamma$, where $X_\gamma$ contains a
prompt photon, the $K$ factor was found to decrease fast with increasing
value of $p_T$, falling below 01 for $p_T\agt14$~GeV~\cite{nlog}.

\begin{theacknowledgments}
The author thanks M. Klasen, L.N. Mihaila, and M. Steinhauser for their
collaboration.
This work was supported in part by BMBF Grant No.\ 05~HT1GUA/4.
\end{theacknowledgments}

\bibliographystyle{aipproc}   

\end{document}